\documentclass[preprint,12pt]{elsarticle} 
\usepackage{graphicx} 
\usepackage{amssymb} 
\usepackage{amsmath} 
\usepackage{amsfonts} 
\usepackage{slashed} 
\usepackage[dvipsnames]{xcolor} 

\newcommand{\nopieft}{\mbox{$\slashed{\pi}$EFT}~} 
\journal{Nuclear Physics A} 

\begin{document} 

\begin{frontmatter} 

\title{Onset of $\eta$-meson binding in the He isotopes} 
\author{N.~Barnea} 
\author{E.~Friedman} 
\author{A.~Gal\corref{cor1}} 
\cortext[cor1]{corresponding author: Avraham Gal, avragal@savion.huji.ac.il}   
\address{Racah Institute of Physics, The Hebrew University, 91904 
Jerusalem, Israel} 

\begin{abstract} 
The onset of binding $\eta$(548) mesons in nuclei is studied in the He 
isotopes by doing precise $\eta NNN$ and $\eta NNNN$ few-body stochastic 
variational method calculations for two semi-realistic $NN$ potentials and two 
energy dependent $\eta N$ potentials derived from coupled-channel models of 
the $N^{\ast}(1535)$ nucleon resonance. The energy dependence of the $\eta N$ 
subthreshold input is treated self consistently. It is found that a minimal 
value of the real part of the $\eta N$ scattering length $a_{\eta N}$ close 
to 1~fm is required to bind $\eta$ mesons in $^3$He, yielding then a few MeV 
$\eta$ binding in $^4$He. The onset of $\eta$-meson binding in $^4$He requires 
that Re$\,a_{\eta N}$ exceeds 0.7~fm approximately. These results compare well 
with results of recent $\eta NNN$ and $\eta NNNN$ pionless effective field 
theory calculations. Related optical-model calculations are also discussed. 
\end{abstract} 


\end{frontmatter}

\section{Introduction} 
\label{sec:intro} 

The near-threshold $\eta N$ system, where $E_{\rm th}(\eta N)=1487$~MeV, 
couples strongly to the nearby $N^{\ast}$(1535) resonance, resulting in 
a fairly attractive and weakly absorptive energy dependent $s$-wave $\eta N$ 
interaction. This was realized, first by coupling the $\eta N$ and $\pi N$ 
channels~\cite{BLi85}, and subsequently verified in works that generate 
dynamically the $N^{\ast}$(1535) $S_{11}$ resonance, e.g. Ref.~\cite{KWW97}, 
by considering the entire set of pseudoscalar meson--octet baryon coupled 
channels. It was soon suggested that $\eta$-nuclear quasibound states might 
exist~\cite{HLi86,LHa86}. While various optical model calculations produce 
invariably such states across the periodic table, the value of mass number 
$A$ that marks the onset of binding depends on which underlying $\eta N$ 
interaction is chosen and on how its subthreshold energy dependence is 
handled~\cite{HLi02,GRI02,JNH02,FGM13,CFG14}. For recent reviews, 
see Refs.~\cite{Gal14,Mares16}. Yet, no such states have ever been 
established beyond doubt~\cite{Wilkin16}. 

Electromagnetic and hadronic production reactions on nuclear targets provide 
useful constraints on the existence of $\eta$ quasibound states in very light 
nuclei. The most recent interpretation of these data is that $\eta d$ is 
unbound, $\eta\,^3$He is nearly or just bound, and $\eta\,^4$He is bound 
\cite{KWi15}. Unfortunately, the $\eta$-nucleus optical model approach 
mentioned above is not trustable in these light nuclei, and genuine few-body 
calculations are required. Our previous few-body $\eta NN$ and $\eta NNN$ 
calculations agree with this conjecture as far as the $\eta d$ and $\eta
\,^3$He systems are concerned~\cite{BFG15}. A similar conclusion for $\eta
\,^3$He has been reached recently by evaluating the $pd\to \eta\,^3$He 
near-threshold reaction~\cite{XLO16}. As for a possible $\eta\,^4$He bound 
state, it has been searched upon with the WASA-at-COSY facility in the $dd\to
\,{^3{\rm He}}N\pi$ reaction~\cite{AAB16}, placing upper limits of a few nb 
on the production of a bound $\eta\,^4$He. 

On the theoretical side, precise $\eta NNN$ and $\eta NNNN$ stochastic 
variational method (SVM) bound-state calculations that use a \nopieft 
(pionless effective field theory) approach have just been published, 
coauthored by us~\cite{BBFG17}. These calculations suggest that the onset 
of $\eta$-meson binding in the $^3$He nucleus requires that the real part 
of the $\eta N$ scattering length $a_{\eta N}$, Re$\,a_{\eta N}$, exceeds 
1~fm approximately, yielding then a few MeV $\eta$ binding in $^4$He, 
and that the onset of $\eta$-meson binding in the $^4$He nucleus requires 
that Re$\,a_{\eta N}$ exceeds 0.7~fm approximately. Another very recent 
work~\cite{Fix17} reports on few-body calculations of the $\eta$-$^3$He and 
$\eta$-$^4$He scattering lengths, concluding that for Re$\,a_{\eta N}\approx 
0.9$~fm neither $\eta\,^3$He nor $\eta\,^4$He are bound. 

Here we extend the SVM few-body bound-state calculations of 
Ref.~\cite{BBFG17}, replacing the \nopieft $NN$ and $NNN$ contact interactions 
used there by the same Minnesota~\cite{MNC77} and Argonne AV4'~\cite{AV402} 
semi-realistic $NN$ interactions used in our previous $\eta NN$ and $\eta NNN$ 
hyperspherical-basis calculations \cite{BFG15}. We confirm the results and 
conclusions of that work for $\eta NN$ and $\eta NNN$. Our results compare 
well with those reached in the \nopieft approach ~\cite{BBFG17}. 

The paper is organized as follows. The two-body $NN$ and $\eta N$ input 
interactions are specified in Sect.~\ref{sec:input} and the way we handle 
the subthreshold energy dependence of the input $\eta N$ interaction 
is described in Sect.~\ref{sec:E}. Results of our SVM calculations of 
$\eta NNN$ and $\eta NNNN$ quasibound states are given in Sect.~\ref{sec:res} 
and compared in Sect.~\ref{sec:OM} with results of optical-model calculations. 
Finally, conclusions are drawn in Sect.~\ref{sec:concl}.

\section{Two-body interaction input} 
\label{sec:input} 

In this section we describe the choice of $NN$ and $\eta N$ pairwise 
interactions. 

\subsection{$NN$ potentials} 
\label{sec:NN} 

Two forms of central $NN$ potentials were used in the present work. 
These are denoted MNC for the Minnesota potential~\cite{MNC77} and AV4p for 
the Argonne AV4' potential~\cite{AV402} parametrized in terms of Gaussians. 
These central potentials were also used in our previous work \cite{BFG15}. 
Both potentials produce nearly identical values of binding energy for 
the weakly bound deuteron, as seen in Table~\ref{tab:Bnuc}, but differ in 
the $^1S_0$ $NN$ low-energy scattering parameters input which enters the 
calculation of the $A$=3,4 nuclei binding energies listed below. They also 
differ in their short-range repulsion which is much stronger in AV4p than 
in MNC. We note that the binding energy of $^3$H within the present AV4' 
parametrization is smaller by 0.11~MeV than that used in our previous 
work~\cite{BFG15}. The proton-proton Coulomb interaction is included in these 
calculations. 

\begin{table}[htb] 
\begin{center} 
\caption{Binding energies $B$ (in MeV) of $s$-shell nuclei calculated 
by applying the SVM to the MNC~\cite{MNC77} and AV4p~\cite{AV402} $NN$ central 
potentials.} 
\begin{tabular}{cccc} 
\hline\hline 
$NN$ int. & $B(^2$H) & $B(^3$H) & $B(^4$He) \\ 
\hline 
MNC  & 2.202 & 8.380 & 29.90 \\ 
AV4p & 2.199 & 8.884 & 32.03 \\ 
Exp. & 2.225 & 8.482 & 28.30 \\  
\hline\hline
\end{tabular} 
\label{tab:Bnuc} 
\end{center} 
\end{table}

\subsection{$\eta N$ potentials} 
\label{sec:etaN} 

\begin{figure}[htb] 
\begin{center} 
\includegraphics[width=0.48\textwidth]{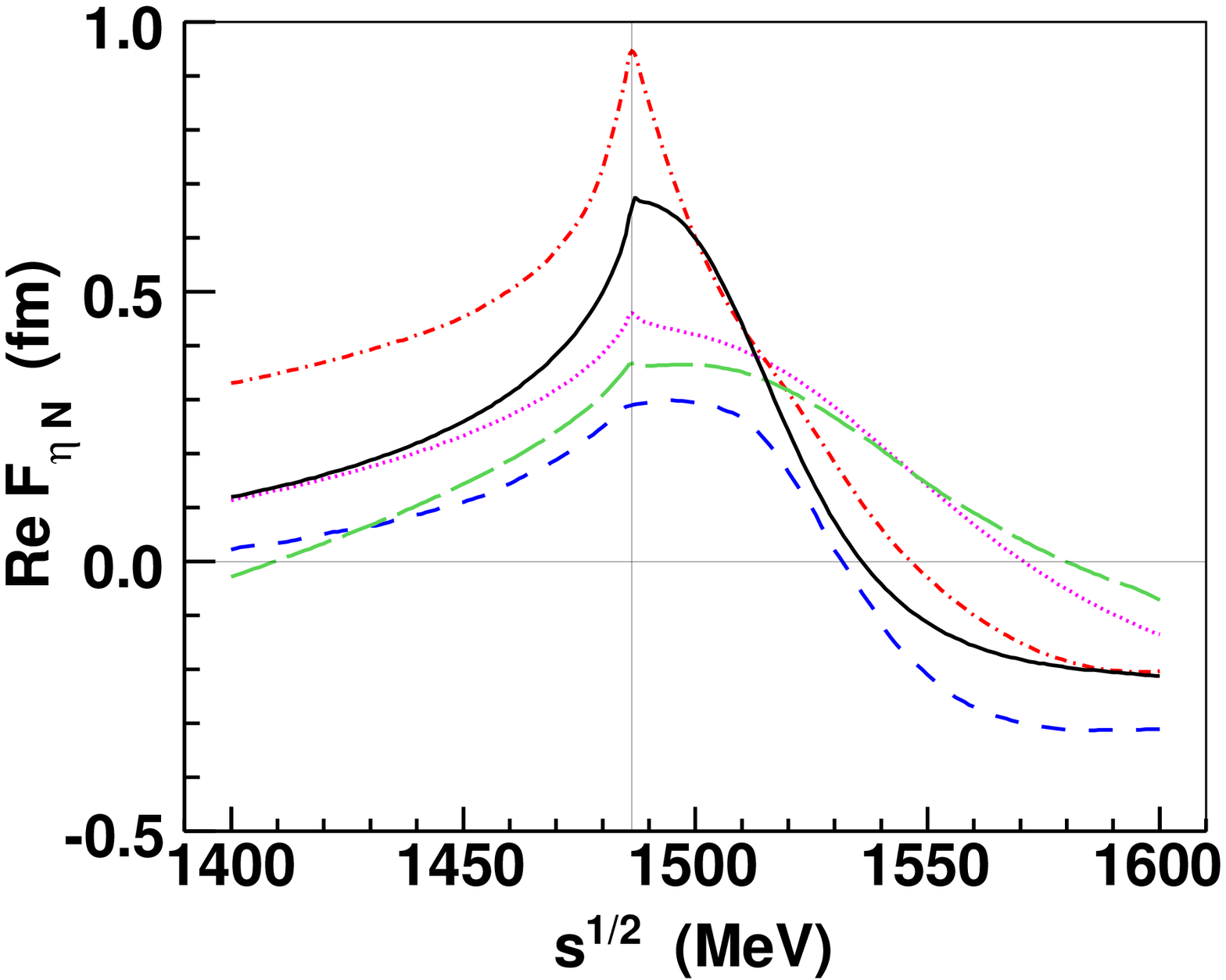} 
\includegraphics[width=0.48\textwidth]{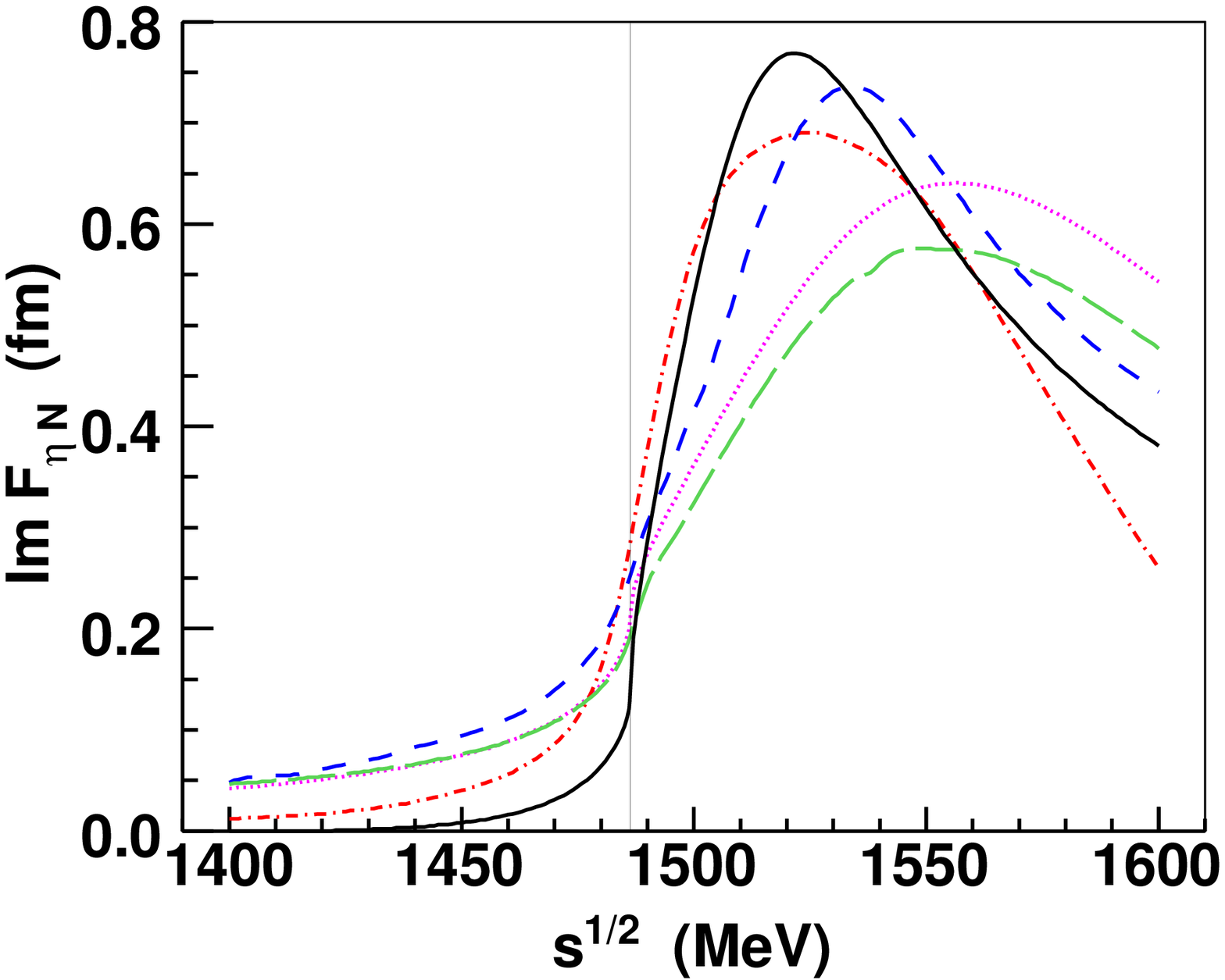} 
\caption{Real (left panel) and imaginary (right panel) parts of the $\eta N$ 
cm $s$-wave scattering amplitude $F_{\eta N}$ as a function of the total 
cm energy $s^{1/2}$ in five meson-baryon coupled-channel interaction models, 
in decreasing order of Re$\,a_{\eta N}$. Dot-dashed curves: GW \cite{GW05}; 
solid: CS \cite{CS13}; dotted: KSW \cite{KSW95}; long-dashed: M2 \cite{MBM12}; 
short-dashed: IOV \cite{IOV02}. The thin vertical line denotes the $\eta N$ 
threshold. Figure adapted from Ref.~\cite{Gal14}.} 
\label{fig:aEtaN1} 
\end{center} 
\end{figure} 

The $\eta N$ interaction has been studied in coupled-channel models that fit 
or generate dynamically the $N^{\ast}(1535)$ $S_{11}$ resonance which peaks 
about 50~MeV above the $\eta N$ threshold. The resulting $\eta N$ scattering 
amplitudes exhibit substantial model dependence, as demonstrated in 
Fig.~\ref{fig:aEtaN1} where the real and imaginary parts of the $\eta N$ 
center-of-mass (cm) $s$-wave scattering amplitude $F_{\eta N}$ are 
plotted as a function of the cm energy $\sqrt{s}\,$ for several coupled 
channel models. A feature in common to all these models is that both real 
and imaginary parts of $F_{\eta N}$ decrease monotonically upon going deeper 
into the subthreshold region. Focusing on the $\eta N$ scattering length 
$a_{\eta N}$, which stands for the value the $\eta N$ scattering amplitude 
$F_{\eta N}$ at threshold, the figure exhibits a wide range of values 
for the real part Re$\,a_{\eta N}$ from 0.2~fm~\cite{KWW97} to almost 
1.0~fm~\cite{GW05}. The imaginary part Im$\,a_{\eta N}$, in contrast, 
is constrained by near-threshold data that involve mostly the coupling 
to the $\pi N$ channel and hence displays a considerably narrower range 
of values, from 0.2 to 0.3~fm. 

Our $\eta N$ potentials are derived from $\eta N$ $s$-wave scattering 
amplitudes $F_{\eta N}$ calculated in two of the meson-baryon coupled-channel 
interaction models, GW~\cite{GW05} and CS~\cite{CS13}, shown in 
Fig.~\ref{fig:aEtaN1}. Whereas the GW model which was considered in our 
previous work~\cite{BFG15} is a $K$-matrix model that accounts for the 
coupling between the $\eta N$ and $\pi N$ channels, the CS model is a genuine  
meson-baryon multi-channel model. The scattering length $a_{\eta N}$ is given 
in these two models by 
\begin{equation} 
a_{\eta N}^{\rm GW}=(0.96+i0.26)~{\rm fm}, \,\,\,\,\,\, \,\,\,\,\,\, 
a_{\eta N}^{\rm CS}=(0.67+i0.20)~{\rm fm}. 
\label{eq:a} 
\end{equation} 

\begin{figure}[thb] 
\begin{center} 
\includegraphics[width=0.46\textwidth]{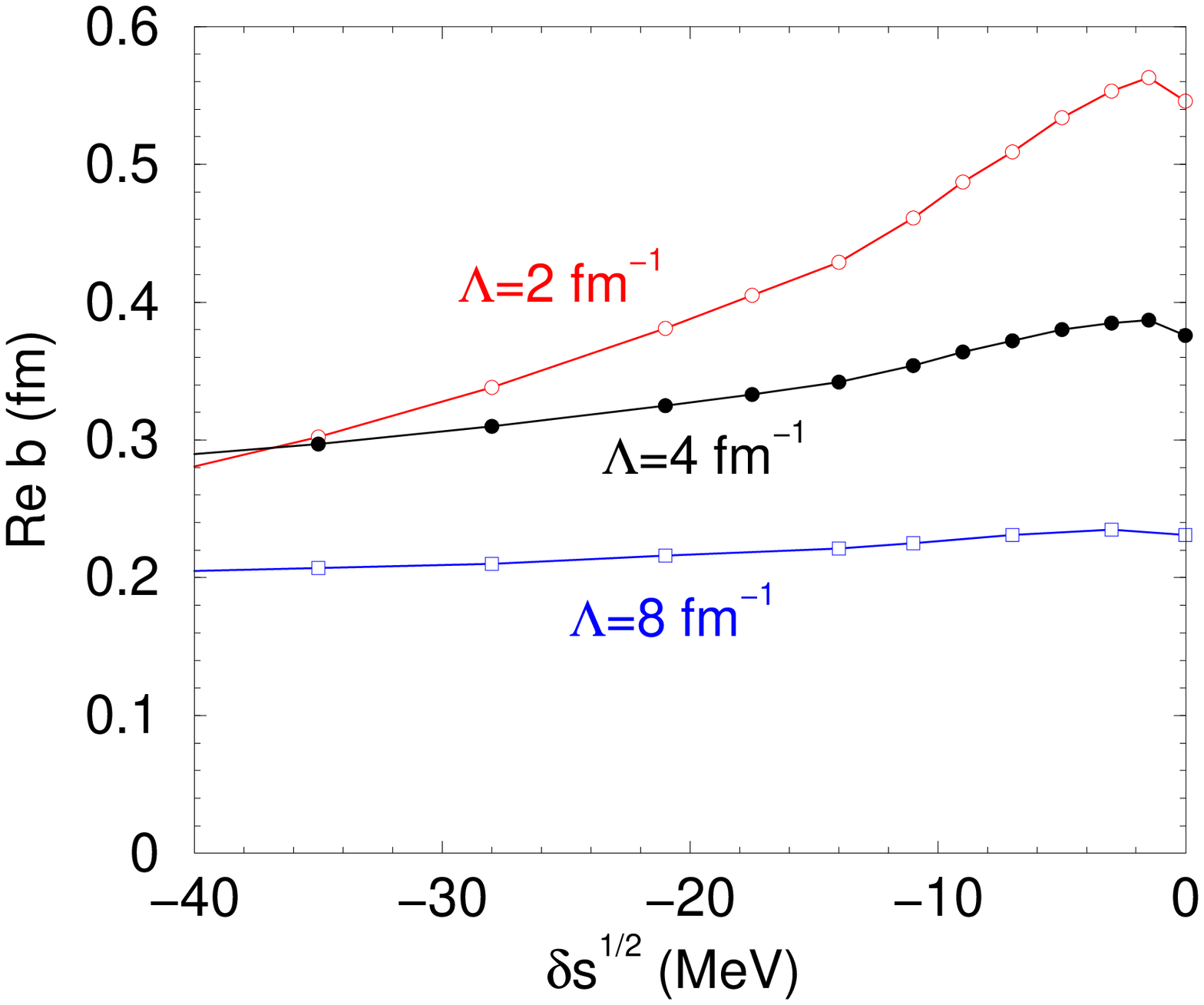} 
\includegraphics[width=0.48\textwidth]{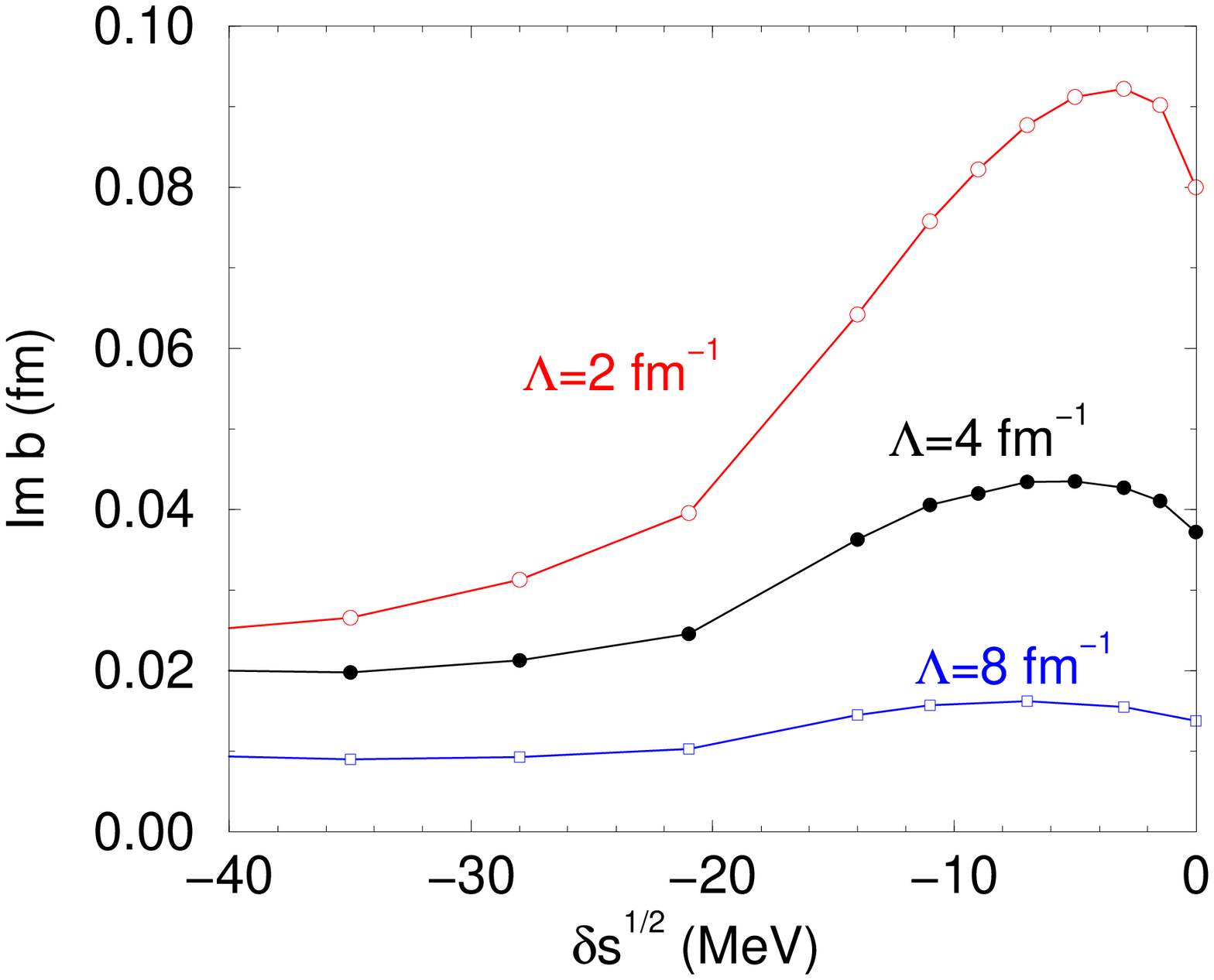} 
\caption{Real (left) and imaginary (right) parts of the strength function 
$b$ of the $\eta N$ effective potential (\ref{eq:v(E)}) at subthreshold 
energies, $\delta\sqrt{s} < 0$, for three values of the scale parameter 
$\Lambda$, as obtained from the scattering amplitude $F_{\eta N}^{\rm GW}$ 
\cite{GW05} shown in Fig.~\ref{fig:aEtaN1}.} 
\label{fig:Wycfit8} 
\end{center} 
\end{figure} 

Following a procedure introduced by Hyodo and Weise~\cite{HW08} for using 
effective $\bar K N$ potentials below threshold, we construct local but 
energy-dependent complex potentials $v_{\eta N}$ that generate the 
$\eta N$ $s$-wave scattering amplitude $F_{\eta N}$ below threshold in 
models GW and CS. This proved useful in $K^-$ few-nucleon calculations 
\cite{DHW08,DHW09,BGL12} and in our previous work on $\eta$ few-nucleon 
quasibound states~\cite{BFG15}. Below we denote by $\delta\sqrt{s}\equiv 
\sqrt{s}-\sqrt{s}_{\rm th}$ the energy argument of $v_{\eta N}$ and of 
$F_{\eta N}$ with respect to the $\eta N$ threshold, with $\delta\sqrt{s}<0$ 
for subthreshold energies. The form chosen for the $\eta N$ potential 
$v_{\eta N}$ is  
\begin{equation}
v_{\eta N}=-\frac{4\pi}{2\mu_{\eta N}}\,b(\delta\sqrt{s})\,\rho_{\Lambda}(r), 
\label{eq:v(E)} 
\end{equation} 
where $\mu_{\eta N}$=346 MeV is the reduced $\eta N$ mass, $b$ is an 
energy dependent strength function, and $\rho_{\Lambda}(r)$ is a normalized 
Gaussian: 
\begin{equation} 
\rho_{\Lambda}(r)=\left(\frac{\Lambda}{2\sqrt{\pi}}\right)^3\,
\exp \left(-\frac{\Lambda^2 r^2}{4}\right), 
\label{eq:rho} 
\end{equation} 
with a scale parameter $\Lambda$ inversely proportional to the range 
of $v_{\eta N}$. As argued in Ref.~\cite{HW08}, $\Lambda$ is related to the 
EFT momentum breakdown scale corresponding to vector-meson exchange: $\Lambda
\lesssim m_{\rho}=3.9$~fm$^{-1}$. A more restrictive upper bound value of 
$\Lambda\lesssim 3.0$~fm$^{-1}$ arises from excluding the $\rho N$ channel 
in dynamically generating the $N^{\ast}(1535)$ resonance~\cite{BFG15}. 
Nevertheless, in order to study the scale dependence of our results, 
we temporarily disregard such constraints on $\Lambda$ and consider 
here three representative values: $\Lambda$=2,4,8~fm$^{-1}$, the latter 
corresponding to extremely short-ranged interaction.  

For a given value of the scale parameter $\Lambda$, the subthreshold values 
of the complex strength function $b$ of Eq.~(\ref{eq:v(E)}) were fitted, as 
detailed in Sect.~2.1 of Ref.~\cite{BFG15}, to the complex phase shifts 
derived from the corresponding values of the subthreshold scattering amplitude 
$F_{\eta N}$. Such strength functions are shown in Fig.~\ref{fig:Wycfit8} 
for three values of the scale parameter $\Lambda$ in model GW~\cite{GW05}. 
The curves for $b$ exhibit monotonic decrease below threshold except for 
some $\Lambda$ dependent humps near threshold that reflect, perhaps, 
the threshold cusp of Re$\,F_{\eta N}$ in Fig.~\ref{fig:aEtaN1}. 
Nevertheless, the subthreshold values $\delta\sqrt{s}$ relevant for 
our calculations are all below $-$10~MeV and are unaffected by these humps. 
Similar curves for $b$ are obtained in model CS, with values smaller uniformly 
for both real and imaginary parts than model GW yields, in accordance with 
the relative strength of the generating subthreshold scattering amplitudes 
$F_{\eta N}$ shown in Fig.~\ref{fig:aEtaN1}. Finally we note that the 
decrease of $b$ upon increasing $\Lambda$, say for Re$\,b$ at threshold, 
follows from the $\Lambda$ independence of the generating Re$\,F_{\eta N}$; 
a $\Lambda$ independent $b$ would necessarily lead to $\eta N$ bound 
states for sufficiently large values of $\Lambda$. 

\section{Energy dependence} 
\label{sec:E} 

Here we outline the self consistent procedure adopted in our previous 
applications~\cite{BFG15,BBFG17} for coping with the energy dependent 
$\eta N$ effective potential $v_{\eta N}$ discussed in Sect.~\ref{sec:etaN}. 
One needs to determine the most appropriate {\it input} value $\delta\sqrt{s}$ 
of the subthreshold $\eta N$ energy at which $v_{\eta N}$ should enter the 
$\eta$-nuclear few-body calculation. To this end, following Ref.~\cite{BGL12}, 
an {\it average} $\eta N$ $\sqrt{s}_{\rm av}$ over the $A$ bound nucleons is 
introduced by 
\begin{equation}
A{\sqrt{s}}_{\rm av}={\sum_{i=1}^{A}\sqrt{(E_{\eta}+E_i)^2-({\vec p}_{\eta}+
{\vec p}_i)^2}}. 
\label{eq:sav}
\end{equation}
Expanding about threshold, with $\sqrt{s_{\rm th}}\equiv E_{\rm th}=m_N+m_{
\eta}=1487$~MeV, one obtains 
\begin{equation}
{\sqrt{s}}_{\rm av}\approx \sqrt{s_{\rm th}}-\frac{B}{A}+\frac{A-1}{A}E_{\eta}
-\frac{1}{A}{\sum_{i=1}^{A}({\vec p}_{\eta}+{\vec p}_i)^2}/(2E_{\rm th}),
\label{eq:thresh}
\end{equation} 
where $B$ is the total binding energy of the $\eta$-nuclear system. 
Transforming in the total cm system the momentum dependent part to kinetic 
energies and taking expectation values in the $\eta$-nuclear bound state, 
one obtains 
\begin{equation} 
\langle\delta\sqrt{s}\rangle = -\frac{B}{A}-\xi_{N}\frac{A-1}{A}\langle 
T_{N:N} \rangle +\frac{A-1}{A}\langle E_{\eta} \rangle -\xi_{\eta}\left ( 
\frac{A-1}{A} \right )^2 \langle T_{\eta} \rangle, 
\label{eq:sqrt{s}1} 
\end{equation} 
where $\xi_{N(\eta)}\equiv m_{N(\eta)}/(m_N+m_{\eta})$. Here, $T_{\eta}$ is 
the $\eta$ kinetic energy operator in the $\eta$-nucleus cm frame, $T_{N:N}$ 
is the pairwise $NN$ kinetic energy operator in the $NN$ pair cm frame and 
$E_{\eta}=(H-H_N)$, with the nuclear Hamiltonian $H_N$ evaluated in its own cm 
frame and the total Hamiltonian $H$ evaluated in the $\eta$-nucleus cm frame. 
We note that the imaginary part of the $\eta N$ interaction, discussed below 
in Sect.~\ref{sec:res}, was excluded from the present construction so that 
both $B$ and $\langle E_{\eta}\rangle$ are real, with $B>0$ and $\langle 
E_{\eta}\rangle <0$ for $\eta$-nuclear bound states. Finally, for $A$=1 
the last three terms on the r.h.s. of Eq.~(\ref{eq:sqrt{s}1}) vanish, leading 
to ${\langle\delta\sqrt{s}\rangle}_{\eta N} = -B$ as expected. 

Since all terms on the r.h.s. are negative definite, the two-body energy 
argument $\delta\sqrt{s}\,$ of $v_{\eta N}$ that enters the few-body 
calculation forms a continuous distribution in the subthreshold region 
that is replaced in non-Faddeev few-body calculations~\cite{Gal14} by the 
output expectation value $\langle\delta\sqrt{s}\rangle$. This expression 
was used in our 2015 calculations for $A=2,3$ employing hyperspherical basis 
expansion of $\eta$-nuclear wavefunctions~\cite{BFG15}. Switching to SVM 
correlated Gaussian basis in our 2017 \nopieft calculations~\cite{BBFG17} as 
well as here, we found it more useful to rewrite Eq.~(\ref{eq:sqrt{s}1}) as 
\begin{equation} 
\langle\delta\sqrt{s}\rangle = -\frac{B}{A} -\xi_{N}\frac{1}{A}\langle T_N 
\rangle +\frac{A-1}{A}\langle E_{\eta}\rangle -\xi_A\xi_{\eta}\left ( 
\frac{A-1}{A} \right )^2 \langle T_{\eta} \rangle, 
\label{eq:sqrt{s}2} 
\end{equation} 
where $\xi_A\equiv Am_N/(Am_N+m_{\eta})$. Here, $T_N$ and $T_{\eta}$ are the 
nuclear and $\eta$ kinetic energy operators evaluated in terms of internal 
Jacobi coordinates, with the total intrinsic kinetic energy of the system 
given by $T=T_N+T_{\eta}$. Since $(A-1)\langle T_{N:N}\rangle$ in 
Eq.~(\ref{eq:sqrt{s}1}) equals $\langle T_N\rangle$ here, 
Eq.~(\ref{eq:sqrt{s}2}) coincides with the former equation apart from 
a kinematical factor $\xi_A$ introduced here to make correspondence with the 
$\eta$-nuclear, last Jacobi coordinate with which $T_{\eta}$ is associated. 

Requiring that the output expectation value $\langle\delta\sqrt{s}\rangle$ 
given by Eq.~(\ref{eq:sqrt{s}2}), as derived from the solution of the few-body 
Schroedinger equation, agrees with the input value $\delta\sqrt{s}\,$ for 
$v_{\eta N}$, this equation defines a self-consistency cycle in our few-body 
$\eta$-nuclear calculations. As argued above, the self consistent energy shift 
$\delta\sqrt{s_{\rm sc}}\,$ is negative definite, with size exceeding 
a minimum nonzero value obtained from the first two terms in the limit of 
vanishing $\eta$ binding. Eq.~(\ref{eq:sqrt{s}2}) in the limit $A\gg 1$ 
coincides with the nuclear-matter expression used in Refs.~\cite{FGM13,CFG14} 
for calculating $\eta$-nuclear quasibound states. 

\begin{figure}[htb] 
\begin{center} 
\includegraphics[width=1.0\textwidth]{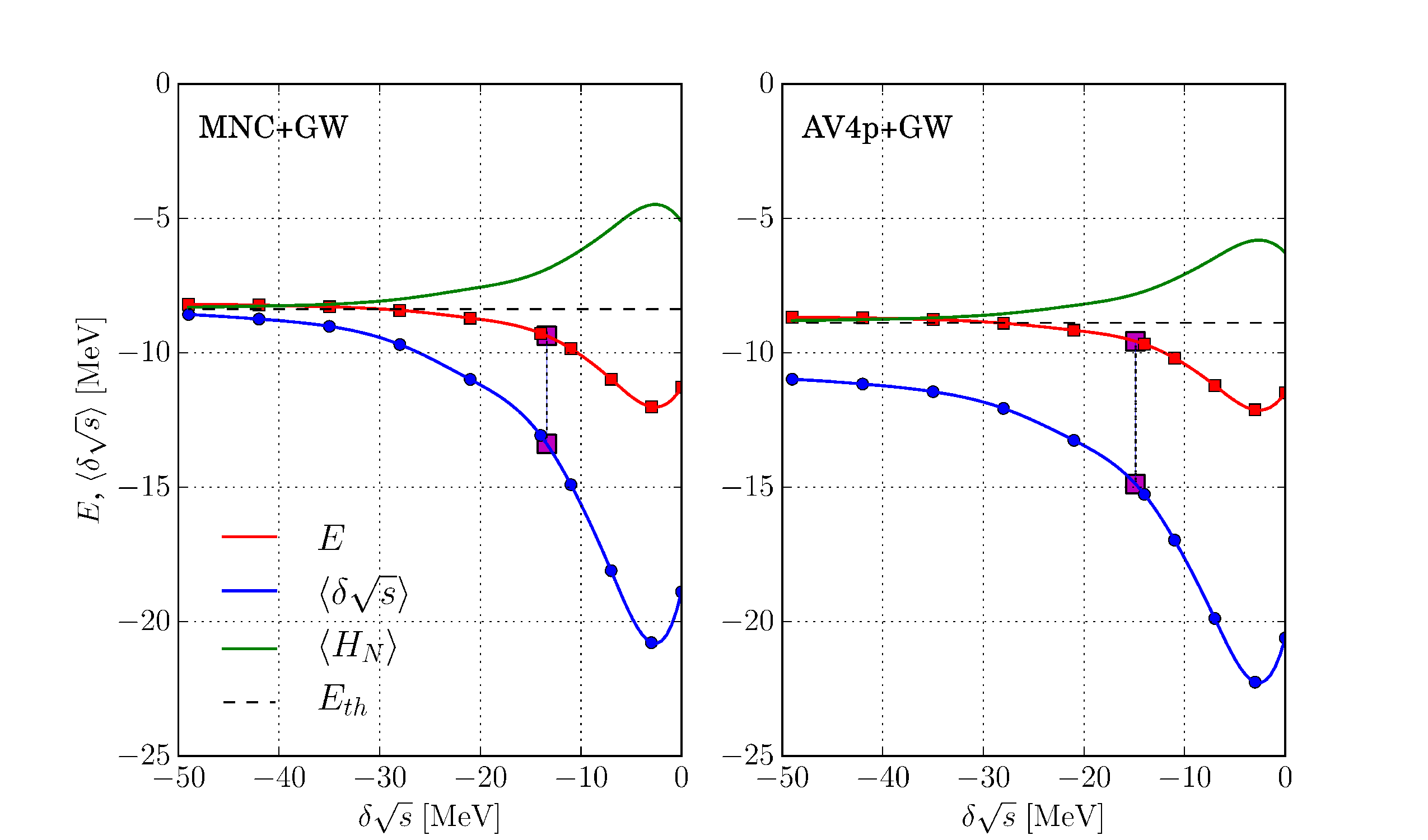} 
\caption{$\eta\,{^3{\rm H}}$ bound-state energies $E$ (squares) and 
expectation values $\langle\delta\sqrt{s}\rangle$ (circles) from 
Eq.~(\ref{eq:sqrt{s}2}), obtained in SVM calculations that use the MNC 
(left panel) and AV4p (right panel) $NN$ potentials, as a function of the 
subthreshold energy input argument $\delta\sqrt{s}\,$ of $v_{\eta N}^{\rm GW}$ 
for $\Lambda=4$~fm$^{-1}$. The dashed vertical line marks the self consistent 
output values of $E$ and $\langle\delta\sqrt{s}\rangle$. The dashed horizontal 
line marks the $^3$H core g.s. energy serving as threshold for a bound $\eta$, 
and the curve above it shows the squeezed core energy $\langle H_N \rangle$.} 
\label{fig:etaT_Lmb4} 
\end{center} 
\end{figure} 

The following two figures demonstrate how the self consistency requirement 
works in actual calculations. The curves plotted in Fig.~\ref{fig:etaT_Lmb4} 
are obtained by interpolating a sequence of calculated $\eta\,^3$H bound-state 
energies (squares) and the corresponding expectation values $\langle\delta
\sqrt{s}\rangle$ (circles) from Eq.~(\ref{eq:sqrt{s}2}) for $A=3$, as a 
function of the input energy argument $\delta\sqrt{s}\,$ of $v_{\eta N}^{
\rm GW}$ with a momentum scale parameter $\Lambda=4$~fm$^{-1}$ and using the 
$NN$ potential MNC (left panel) and AV4p (right panel). The dashed vertical 
line marks the self consistent value of $\delta\sqrt{s}\,$ at which the 
outcome bound-state energy $E(\eta\,{^3{\rm H}})$ is evaluated. Both lower and 
middle curves are located, as they should, {\it below} the dashed horizontal 
line that marks the $^3$H core bound-state energy which serves as threshold 
energy for the $\eta\,{^3{\rm H}}$ bound state. In contrast, the upper curve 
which shows the expectation value $\langle H_N \rangle$ of the nuclear core 
energy is located {\it above} this dashed horizontal line. Finally, we note 
that the $\eta$ binding (or more precisely separation) energy $B_{\eta}$ is 
about 1.0~MeV in the left panel (MNC) and 0.7~MeV in the right one 
(AV4p), a few MeV less than what a $v_{\eta N}$ fitted exclusively to 
$F_{\eta N}$ at threshold yields. The somewhat weaker $\eta$ binding for 
AV4p with respect to that for MNC is related to the considerably stronger 
short-range repulsive component present in the $NN$ potential AV4p. 
Although computed for $\eta\,^3$H, these $B_{\eta}$ values agree to 
within 15$\pm$10~keV with those that were checked in a sample $\eta\,^3$He 
calculation, so from here on we shall treat them as $\eta\,^3$He results. 
These $B_{\eta}$ values are also consistent with those calculated by us 
previously~\cite{BFG15}. 

\begin{figure}[!ht] 
\begin{center} 
\includegraphics[width=0.7\textwidth]{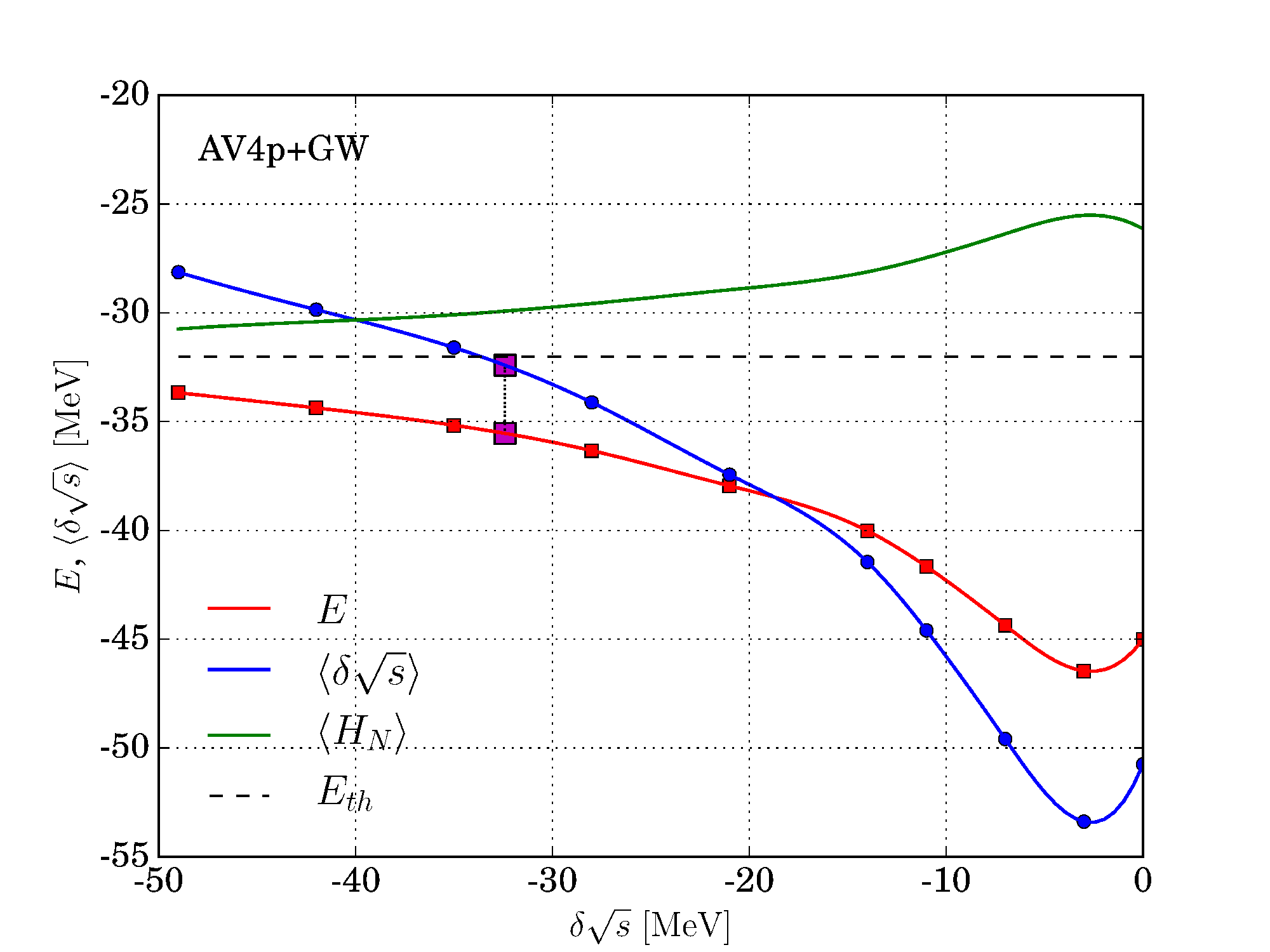}
\caption{Same as in Fig.~\ref{fig:etaT_Lmb4}, but for $\eta\,{^4{\rm He}}$ 
using the AV4p $NN$ potential.} 
\label{fig:eta4He_Wyc_Lmb4} 
\end{center} 
\end{figure} 

Fig.~\ref{fig:eta4He_Wyc_Lmb4} for $\eta\,{^4{\rm He}}$ is similar to 
Fig.~\ref{fig:etaT_Lmb4} (right) for $\eta\,{^3{\rm H}}$, using the AV4p $NN$ 
potential. The self consistent value of $B_{\eta}$, about 3.5~MeV, is larger 
here than the 0.7~MeV there; however it is much smaller, almost by 10~MeV, 
than what a threshold $v_{\eta N}$ yields, owing to the subthreshold 
energy dependence of $v_{\eta N}$. Note that the self consistent value 
$\delta\sqrt{s_{\rm sc}}\,$ in the compact $^4$He core (about $-$32~MeV, 
see Fig.~\ref{fig:eta4He_Wyc_Lmb4}) is considerably farther away from 
threshold than its counterpart in $^3$H 
(about $-$15~MeV, see Fig.~\ref{fig:etaT_Lmb4}, right).

\section{Results} 
\label{sec:res} 

\begin{figure}[!ht] 
\begin{center} 
\includegraphics[width=0.7\textwidth]{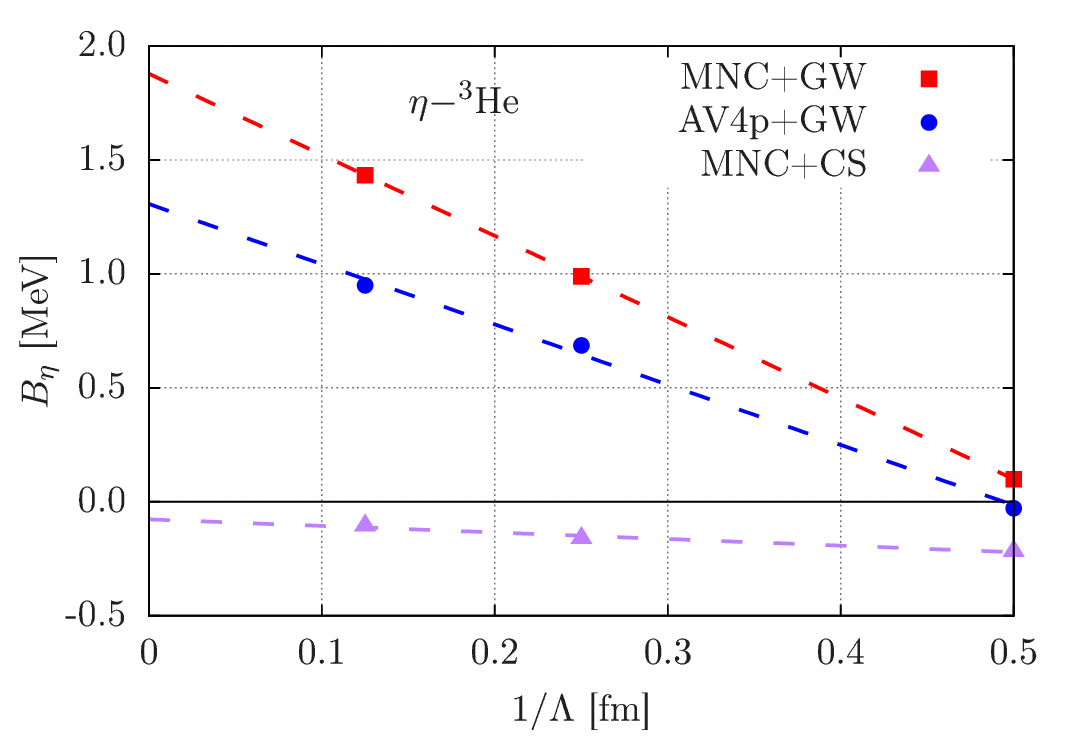} 
\caption{$B_{\eta}(\eta\,{^3{\rm He}})$ values from Table~\ref{tab:3He} 
as a function of $1/\Lambda$, calculated within several combinations of 
($NN,\,\eta N$) potentials marked in the upper-right corner.} 
\label{fig:E_etaT} 
\end{center} 
\end{figure}

\begin{figure}[!ht] 
\begin{center} 
\includegraphics[width=0.7\textwidth]{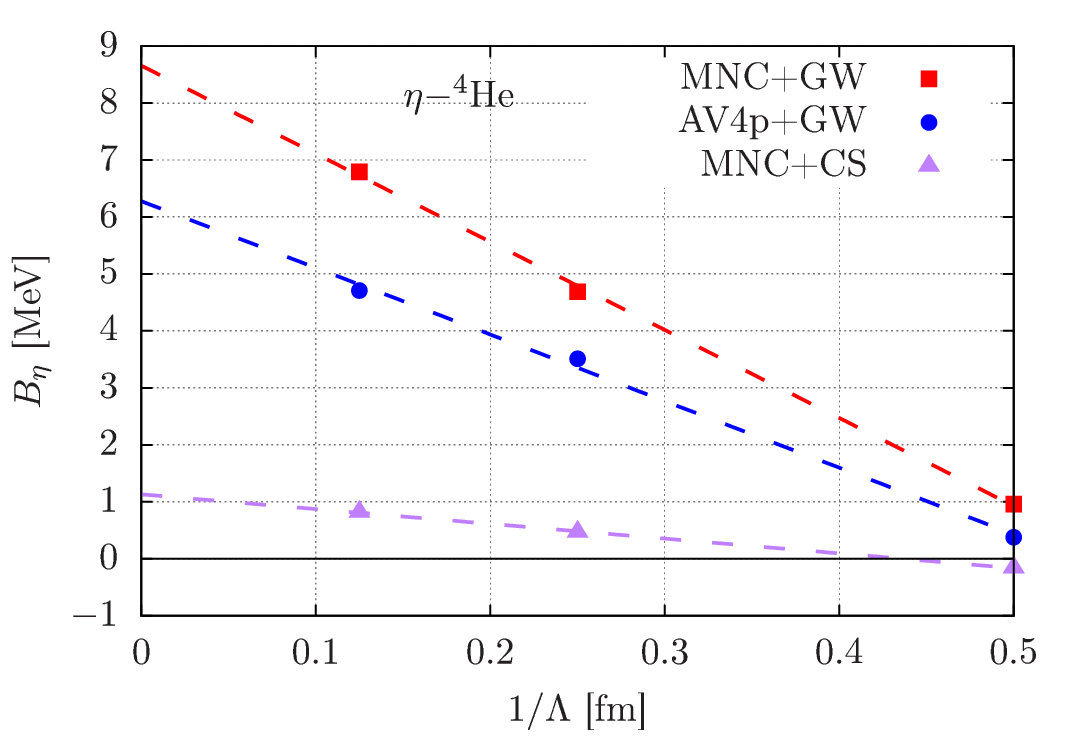}
\caption{$B_{\eta}(\eta\,{^4{\rm He}})$ values from Table~\ref{tab:4He} 
as a function of $1/\Lambda$, calculated within several combinations of 
($NN,\,\eta N$) potentials marked in the upper-right corner.} 
\label{fig:E_eta4He} 
\end{center} 
\end{figure}

The present few-body self consistent stochastic-variational-method (SVM) 
calculations follow those of Ref.~\cite{BBFG17} where this method is briefly 
discussed. Our results for $\eta\,{^3{\rm He}}$ bound states are listed in 
Table~\ref{tab:3He} and shown in Fig.~\ref{fig:E_etaT}, and for $\eta\,{^4{
\rm He}}$ in Table~\ref{tab:4He} and Fig.~\ref{fig:E_eta4He}, for several 
choices of ($NN,\eta N$) potential combinations, each one for three 
values of the momentum scale parameter $\Lambda$. Along with $\eta$ binding 
(separation) energies $B_{\eta}$ and widths $\Gamma_{\eta}$, we also list 
the self consistent values $\delta\sqrt{s_{\rm sc}}\,$ at which these values 
of $B_{\eta}$ and $\Gamma_{\eta}$ were evaluated, plus expectation values 
of potential and kinetic energies. Noting that Im$\,b$$\,\ll\,$Re$\,b$, 
see Fig.~\ref{fig:Wycfit8}, the binding energies $B_{\eta}$ were evaluated 
using real Hamiltonians in which Im$\,v_{\eta N}$ was neglected, and the 
$\eta$-nuclear widths $\Gamma_{\eta}$ were calculated perturbatively using 
wavefunctions $\Psi_{\rm g.s.}$ generated by these real Hamiltonians, viz.  
\begin{equation}
\Gamma_{\eta} = -2\, \langle \,\Psi_{\rm g.s.}\, | \, {\rm Im} \, V_{\eta} \, 
| \, \Psi_{\rm g.s.} \, \rangle \;.
\label{eq:Gamma} 
\end{equation} 
Here, $V_{\eta}$ sums over all pairwise $\eta N$ interactions. Since 
$|{\rm Im}\,V_{\eta}|\ll |{\rm Re}\,V_{\eta}|$, this is a reasonable 
approximation. Restoring Im$\,V_{\eta}$, it was estimated, by solving 
the corresponding optical-model $\eta$-nucleus bound-state equation with 
and without Im$\,V_{\eta}$, that $B_{\eta}$ decreases by less than 0.3~MeV 
for weakly bound states. 

\begin{table}[htb]
\begin{center}
\caption{$\eta\,^3$He bound state energies, widths and shifts (in MeV) from SVM 
calculations using 3 values of $\Lambda$ (in fm$^{-1}$) for several choices 
of ($NN,\,\eta N$) pairwise potentials. The listed values of $\Gamma_{\eta}$ 
outdate the excessively large widths listed in Ref.~\cite{BFG15} that arose 
from a programming error. The number of displayed digits reflects the 
numerical accuracy of these calculations.} 
\begin{tabular}{cccccccc} 
\hline \hline
$\Lambda$ & $\delta\sqrt{s_{\rm sc}}$ & $\langle V_N \rangle$ & $\langle 
T_N \rangle$ & $\langle V_{\eta} \rangle$ & $\langle T_{\eta} \rangle$ & 
$B_{\eta}$ & $\Gamma_{\eta}$ \\ 
\hline
\multicolumn{8}{c}{$NN$: MNC~~~~~~~$\eta N$: GW} \\ 
2 & $-$9.385 & $-$36.636 & 28.384 & $-$3.376 &  3.148 & 0.099 & 1.144 \\ 
4 & $-$13.392 &$-$39.655 & 32.767 &$-$15.002 & 12.520 & 0.990 & 3.252 \\ 
8 & $-$18.787 &$-$40.983 & 37.534 &$-$30.956 & 24.592 & 1.433 & 3.280 \\ 
\hline
\multicolumn{8}{c}{$NN$: AV4p~~~~~~~$\eta N$: GW}  \\ 
2 & $-$11.478&$-$47.599 & 38.805 & $-$2.365 &  2.304 & $-$0.028 & 0.769 \\ 
4 & $-$14.881&$-$51.199 & 43.383 &$-$11.885 & 10.131 &    0.686 & 2.438 \\ 
8 & $-$18.234&$-$52.182 & 46.438 &$-$21.981 & 17.891 &    0.950 & 2.332 \\ 
\hline
\multicolumn{8}{c}{$NN$: MNC~~~~~~~$\eta N$: CS}  \\ 
2 & $-$8.388 & $-$35.600 & 27.243 & $-$0.251 & 0.446 & $-$0.217 & 0.057 \\ 
4 & $-$8.712 & $-$35.931 & 27.657 & $-$1.312 & 1.367 & $-$0.161 & 0.227 \\ 
8 & $-$9.402 & $-$36.263 & 28.352 & $-$3.509 & 3.145 & $-$0.105 & 0.385 \\ 
\hline\hline
\end{tabular}
\label{tab:3He}
\end{center}
\end{table}

The tables demonstrate that the smaller the range ($\sim 1/\Lambda$) of the 
$\eta N$ interaction, the larger is the resulting $\eta$ binding energy $B_{
\eta}$, in spite of the increased value of $-\delta\sqrt{s_{\rm sc}}$ which 
implies a weaker $\eta N$ potential strength $b$. 
In fact, all of the entities listed in these tables increase in magnitude 
with $\Lambda$, particularly $\langle V_{\eta} \rangle$ and $\langle T_{\eta} 
\rangle$ once the $\eta$-nuclear system becomes bound. Nevertheless, the 
$\eta$ averaged momentum encountered there reaches a modest maximal value of 
$\langle p^2_{\eta}\rangle ^{\frac{1}{2}}=1.08$~fm$^{-1}$, for (MNC, GW) with 
$\Lambda=8$~fm$^{-1}$ in $\eta\,^4$He, in spite of the much larger momenta 
that this scale parameter is capable of generating. For a given $\eta N$ 
potential, here GW, more binding is obtained with the MNC $NN$ potential 
than with the more realistic AV4p, as already noted in the discussion of 
Fig.~\ref{fig:etaT_Lmb4}. And finally, for a given $NN$ potential, here MNC, 
more binding is obtained obviously using GW than the weaker CS for the $\eta 
N$ potential. In fact CS does not bind $\eta\,^3$He, as indicated by the 
negative values for $B_{\eta}$ listed in the last group in Table~\ref{tab:3He} 
and as shown by the lower curve in Fig.~\ref{fig:E_etaT}. These results 
are consistent with those in Ref.~\cite{BFG15}. For $\eta\,^4$He, although 
Table~\ref{tab:4He} and Fig.~\ref{fig:E_eta4He} suggest that a bound state 
exists for the combination ($NN$:MNC, $\eta N$:CS), it is doubtful whether 
CS can really bind $\eta\,^4$He for the more realistic AV4p $NN$ interaction. 

\begin{table}[htb]
\begin{center}
\caption{SVM calculations of $\eta\,{^4{\rm He}}$ bound states, 
see caption of Table~\ref{tab:3He} for details.}
\begin{tabular}{cccccccc}
\hline\hline
$\Lambda$ & $\delta\sqrt{s_{\rm sc}}$ & $\langle V_N \rangle$ & $\langle 
T_N \rangle$ & $\langle V_{\eta} \rangle$ & $\langle T_{\eta} \rangle$ & 
$B_{\eta}$ & $\Gamma_{\eta}$ \\
\hline
\multicolumn{8}{c}{$NN$: MNC~~~~~~~$\eta N$: GW} \\ 
2 & $-$19.477& $-$90.052& 60.330& $-$8.783 &  7.645 & 0.96 & 1.975  \\
4 & $-$29.750& $-$95.510& 68.527& $-$32.925& 25.320 & 4.69 & 4.500  \\
8 & $-$43.294& $-$97.754& 78.466& $-$65.387& 47.982 & 6.79 & 6.196  \\
\hline
\multicolumn{8}{c}{$NN$: AV4p~~~~~~~$\eta N$: GW}  \\ 
2 & $-$23.646& $-$120.760& 88.965& $-$6.356 &  5.744 & 0.38 & 1.207  \\
4 & $-$32.411& $-$127.922& 97.998& $-$26.851& 21.233 & 3.51 & 3.615 \\
8 & $-$40.279& $-$129.720& 104.085&$-$46.116& 34.994 & 4.71 & 4.198 \\
\hline
\multicolumn{8}{c}{$NN$: MNC~~~~~~~$\eta N$: CS}  \\ 
2 & $-$16.704& $-$88.149& 58.239& $-$0.938 &  1.107 & $-$0.16 & 0.133 \\
4 & $-$19.246& $-$89.948& 60.566& $-$8.468 &  7.483 &    0.47 & 0.901  \\
8 & $-$22.434& $-$90.942& 63.210& $-$17.007& 14.021 &    0.82 & 1.108  \\
\hline\hline
\end{tabular}
\label{tab:4He}
\end{center}
\end{table} 

Figs.~\ref{fig:E_etaT} and \ref{fig:E_eta4He} suggest a significant 
theoretical uncertainty in the computed $\eta$ binding energies $B_{\eta}$ 
for two reasons. First, for a given choice of $NN$ and $\eta N$ interaction 
models, say the (AV4p, GW) combination, the computed values of $B_{\eta}$ 
depend on the scale parameter $\Lambda$ of the Gaussian shape assumed for 
$v_{\eta N}$. As stated above, the smaller the range ($\sim 1/\Lambda$) of 
the $\eta N$ interaction, the larger $B_{\eta}$ is, consistently with the 
observation made long ago for $B_{\Lambda}$ in $\Lambda$ hypernuclear few-body 
calculations \cite{GL81}. In our previous work~\cite{BFG15} we argued that 
$\Lambda \lesssim 3$~fm$^{-1}$ holds implicitly in the $N^{\ast}$(1535) 
resonance meson-baryon models within which the $\eta N$ scattering amplitude 
$F_{\eta N}$ is determined. Excluding therefore as high values as 
$\Lambda=8$~fm$^{-1}$, this would mean that the GW $\eta N$ interaction 
binds both He isotopes, with $B_{\eta}(\eta\,{^3{\rm He}})$ about 0.3~MeV 
or less and $B_{\eta}(\eta\,{^4{\rm He}})\lesssim 2$~MeV upon using the 
more realistic AV4p $NN$ interaction. The second origin of theoretical 
uncertainty concerns the choice of $\eta N$ interaction model: choosing CS 
instead of the GW interaction, for example, one could envisage an unbound 
$\eta\,^3$He and a very slightly bound $\eta\,^4$He. Finally, if the $\eta N$ 
interaction is weaker than CS, it is likely that neither $\eta\,^3$He nor 
$\eta\,^4$He are bound. In the next section we discuss, for comparison, 
optical-model calculations of $\eta\,^3$He and $\eta\,^4$He bound states.

\section{Comparison with other model calculations} 
\label{sec:OM} 

In this section we connect between the few-body $\eta$-nuclear calculations 
pursued in the main part of the present work and optical-model approaches 
applied to heavier nuclei, normally for $A\geq 12$, as summarized by Mare\v{s} 
et al.~\cite{Mares16}. In particular, Xie et al.~\cite{XLO16} recently fitted 
the strongly energy dependent $dp\to \eta\,^3$He cross sections near threshold 
using a `$t_{\eta N}\rho_A$' optical model approach, where $t_{\eta N}$ is 
directly related to the $\eta N$ scattering length $a_{\eta N}$ and $\rho_A$ 
is the nuclear g.s. static density normalized in coordinate space to $A$. 
These authors derived an effective value 
\begin{equation} 
a^{\rm eff}_{\eta N}=[(0.48\pm 0.05)+i(0.18\pm 0.02)]~{\rm fm}, 
\label{eq:aeff} 
\end{equation} 
claiming that although insufficient to generate a bound state pole, 
nevertheless it generates a virtual-state pole near threshold corresponding 
to $B_{\eta}\approx -0.3$~MeV and $\Gamma\approx 3.0$~MeV. This effective 
value $a^{\rm eff}_{\eta N}$ is necessarily smaller than the value of the 
scattering length $a_{\eta N}$ owing to the reduction that both values of 
Re$\,F_{\eta N}$ and Im$\,F_{\eta N}$ undergo below threshold. The actual 
value of $a_{\eta N}$ in their model could well come close to that of 
$a_{\eta N}^{\rm CS}$, Eq.~(\ref{eq:a}). It is interesting then to see whether 
the larger values in models GW and CS of Re$\,a_{\eta N}$ compared to the 
`effective' values cited in Eq.~(\ref{eq:aeff}) are able to generate $\eta\,
^3$He and $\eta\,^4$He bound states in this optical model approach. 

Within the underlying Watson (W) multiple scattering series \cite{EK80}, 
the $t\rho$ optical potential in momentum space assumes the form 
\begin{equation} 
{\tilde V}^{\rm W}_{\eta A}(q)=-\frac{4\pi}{2\mu_{\eta N}}\, 
{\cal F}_{\eta N}(q)\,{\tilde\rho}_A(q), 
\label{eq:Vopt} 
\end{equation} 
where the $\eta N$ subthreshold energy dependence is implicit. 
A Gaussian momentum dependence was adopted for both the cm scattering 
amplitude ${\cal F}_{\eta N}$ and the momentum-space nuclear density 
${\tilde\rho}_A$: 
\begin{equation} 
{\cal F}_{\eta N}(q)=F_{\eta N}\exp(-\frac{q^2}{\Lambda^2}), \,\,\,\,\, 
{\tilde\rho}_A(q)=A\exp(-\frac{q^2}{\lambda_A^2}), 
\label{eq:Fgauss} 
\end{equation}  
with $\Lambda$ chosen the same as in the Gaussian form (\ref{eq:v(E)}) 
of $v_{\eta N}$. Fourier transforming the product ${\cal F}_{\eta N}(q)
\times {\tilde\rho}_A(q)$ to coordinate space, we obtain 
\begin{equation} 
\exp(-\frac{q^2}{\Lambda^2}){\tilde\rho}_A(q) \Rightarrow 
A(R_A\sqrt{\pi})^{-3}\exp(-\frac{r^2}{R_A^2}), \,\,\,\,\, R_A^2=r_0^2+r_A^2, 
\label{eq:Gauss} 
\end{equation} 
with $r_0=2/\Lambda$ and $r_A=2/\lambda_A$, where $r_3=1.436$~fm and 
$r_4=1.165$~fm are the Gaussian size parameters of the point-nucleon density 
distributions of $^3$He and $^4$He, respectively, derived by unfolding the 
proton charge distribution from the nuclear charge distribution~\cite{ADN13}. 
Finally, to incorporate the leading $1/A$ correction, we go to the KMT form 
of the optical potential~\cite{KMT59}, as applied e.g. by Feshbach et 
al.~\cite{FGH71}, multiplying the Watson optical potential (\ref{eq:Vopt}) 
by $(A-1)/A$.{\footnote{This $(A-1)/A$ factor takes care of the 
antisymmetrization of nucleons in intermediate multiple-scattering states. To 
get the meson-nuclear overall $T$ matrix, one then multiplies the $T$ matrix 
arising from $V^{\rm KMT}$ by $A/(A-1)$; see Sect.~4.2 in Ref.~\cite{EK80} 
for a detailed exposition.}} Hence: 
\begin{equation} 
V^{\rm KMT}_{\eta A}(r)=-\frac{4\pi}
{2\mu_{\eta N}}\,F_{\eta N}\,(A-1)\,(R_A \sqrt{\pi})^{-3}
\exp(-\frac{r^2}{R_A^2}).  
\label{eq:KMT} 
\end{equation} 

\begin{table}[htb]
\begin{center}
\caption{$1s_{\eta}$ binding energies $B_{\eta}$ and widths $\Gamma_{\eta}$ 
(in MeV) calculated for $\eta\,^3$He and $\eta\,^4$He by using the KMT 
$\eta$-nuclear optical potential (\ref{eq:KMT}) for $\eta N$ scale parameter 
$\Lambda=4$~fm$^{-1}$ and with scattering amplitudes $F^{\rm GW}_{\eta N}$ 
and $F^{\rm CS}_{\eta N}$, each evaluated at threshold $\delta\sqrt{s}=0$ and 
also at self-consistent subthreshold energies $\delta\sqrt{s_{\rm sc}}\,$ 
(in MeV) listed in Sect.~\ref{sec:res}. Results from the present few-body (FB) 
calculations, with AV4p along with GW and MNC along with CS, and from those in 
Ref.~\cite{BBFG17} are also listed.} 
\begin{tabular}{cccccccc} 
\hline \hline
$\eta N$ & $\eta$-nuclear & \multicolumn{3}{c}{$\eta\,^3$He} & 
\multicolumn{3}{c}{$\eta\,^4$He} \\ 
model & model & $\delta\sqrt{s}$ & $B_{\eta}$ & $\Gamma_{\eta}$ & 
$\delta\sqrt{s}$ & $B_{\eta}$ & $\Gamma_{\eta}$ \\
\hline 
GW~\cite{GW05} & optical      & 0 & 0.33 & 6.04 & 0 & 25.1 & 40.8  \\ 
               & FB (present) & 0 & 2.60 & 5.08 & 0 & 13.0 & 12.0  \\ 
     & FB~\cite{BBFG17} & 0 & 3.62 & 7.52 & 0 & 10.8 & 13.2  \\ 
 & & & & & & & \\ 
                & optical & $-$14.9 &  --  &  --  & $-$32.4 & 1.03 & 2.35  \\ 
           & FB (present) & $-$14.9 & 0.69 & 2.44 & $-$32.4 & 3.51 & 3.62  \\ 
       & FB~\cite{BBFG17} & $-$21.1 & 0.30 & 1.46 & $-$32.2 & 1.54 & 2.82  \\ 
 & & & & & & & \\ 
CS~\cite{CS13} & optical      & 0 &  --  &  --  & 0 & 6.39 & 21.0  \\ 
           & FB (present) & 0 & 0.30 & 2.16 & 0 & 6.77 & 11.3  \\ 
       & FB~\cite{BBFG17} & 0 &  --  &  --  & 0 & 3.47 & 8.95  \\       
 & & & & & & & \\ 
                & optical & $-$8.7  &  --  &  --  & $-$19.2 &  --  &  --  \\   
           & FB (present) & $-$8.7  &  --  &  --  & $-$19.2 & 0.47 & 0.90 \\ 
       & FB~\cite{BBFG17} & $-$9.9    &  --  &  --  & $-$23.7 &  --  &  --  \\ 
\hline\hline
\end{tabular}
\label{tab:oset}
\end{center}
\end{table}

In Table~\ref{tab:oset} we list $\eta$ binding energies and widths calculated 
in the He isotopes using the optical potential (\ref{eq:KMT}) for $\eta N$ 
scale parameter $\Lambda=4$~fm$^{-1}$ and with the GW and CS scattering 
amplitudes $F_{\eta N}$ evaluated at both threshold energy $\delta\sqrt{s}=0$, 
and at the subthreshold energies $\delta\sqrt{s_{\rm sc}}\,$ found in the 
few-body self consistent calculations listed in Tables~\ref{tab:3He} 
and~\ref{tab:4He} for ($NN$,$\eta N$) potential combinations (AV4p, GW) 
and (MNC, CS). The results show that of the four cases studied within the 
optical-model approach $\eta\,^3$He is bound, and quite slightly so, only by 
choosing the scattering amplitude $F^{\rm GW}_{\eta N}$ at threshold to work 
with. Once a self consistent value $\delta\sqrt{s_{\rm sc}}\,$ is used to 
represent the $\eta N$ subthreshold energy appropriate in a few-body $\eta 
NNN$ calculation, the slight $\eta\,^3$He binding shown in the first row of 
the table disappears. The $\eta N$ CS scattering amplitude does not produce 
binding at all, in agreement with the $\eta\,^3$He few-body calculations 
listed in Table~\ref{tab:3He}. For $\eta\,^4$He both optical-model strengths 
at threshold produce substantial values of $B_{\eta}$ and particularly of 
$\Gamma_{\eta}$. Here too, consideration of the appropriate subthreshold 
energy reduces these values, leaving $\eta\,^4$He weakly bound only for the 
GW choice of $\eta N$ interaction. Finally, the table also demonstrates 
good agreement between the present SVM results and those obtained in 
recent \nopieft SVM calculations.

\section{Conclusion} 
\label{sec:concl} 

In conclusion, we have presented truly few-body SVM calculations of 
$\eta NNN$ ($\eta\,^3$He) and $\eta NNNN$ ($\eta\,^4$He) bound states, 
using semi-realistic $NN$ interactions and $\eta N$ subthreshold interactions 
derived in coupled channels studies of the $N^{\ast}(1535)$ nucleon resonance. 
Considering $\eta N$ scale parameters $\Lambda=2,4$~fm$^{-1}$, while excluding 
as high value as $\Lambda=8$~fm$^{-1}\gg m_{\rho}$ which corresponds to 
extremely short-ranged interaction, the present results suggest that the onset 
of $\eta\,^3$He binding occurs for Re$\,a_{\eta N}$ close to 1~fm, as in model 
GW~\cite{GW05}, consistently with our previous hyperspherical-basis $\eta NNN$ 
calculations~\cite{BFG15}. The onset of $\eta\,^4$He binding requires a lower 
value of Re$\,a_{\eta N}$ around 0.7~fm which is comfortably satisfied in 
model GW and almost in model CS~\cite{CS13}. These results are also in good 
agreement with the very recent SVM calculations coauthored by us~\cite{BBFG17} 
which use a \nopieft approach. Further dedicated experimental searches for 
$\eta\,^4$He bound states are needed in order to confirm the recent negative 
report from COSY~\cite{AAB16} which, taken at face value, implies that 
Re~$a_{\eta N}\lesssim 0.7$~fm. 

We have also compared our few-body calculations with leading-order optical 
model calculations that use the same subthreshold energies $\delta\sqrt{s}$ 
as those determined self consistently in the few-body calculations. Careful 
attention was paid to $1/A$ corrections. These optical model calculations, 
which produce less binding than that produced in the corresponding few-body 
calculations, nevertheless give {\it qualitatively} similar results to those 
of the few-body calculations. Ignoring the energy dependence of the input 
$\eta N$ amplitude leads to strongly excessive binding energies, and widths, 
in both approaches.

\section*{Acknowledgments} 

We thank Ji\v{r}\'{i} Mare\v{s} and Martin Schaefer for helpful discussions on 
related matters, and Eulogio Oset and Colin Wilkin for helpful exchanges on 
the contents of Ref.~\cite{XLO16}. This work was supported in part (NB) by the 
Israel Science Foundation grant 1308/16 and by PAZI Fund grants, and in part 
(EF, AG) by the EU initiative FP7, Hadron-Physics3, under the SPHERE and 
LEANNIS cooperation programs.


\begin{thebibliography}{99}

\bibitem{BLi85} R.S.~Bhalerao, L.C.~Liu, Phys. Rev. Lett. 54 (1985) 865. 

\bibitem{KWW97} N.~Kaiser, T.~Waas, W.~Weise, Nucl. Phys. A 612 (1997) 297, 
and references to earlier work cited therein. 

\bibitem{HLi86} Q.~Haider, L.C.~Liu, Phys. Lett. B 172 (1986) 257. 

\bibitem{LHa86} L.C.~Liu, Q.~Haider, Phys. Rev. C 34 (1986) 1845. 

\bibitem{HLi02} Q.~Haider, L.C.~Liu, Phys. Rev. C 66 (2002) 045208. 

\bibitem{GRI02} C.~Garc\'{i}a-Recio, T.~Inoue, J.~Nieves, E.~Oset, 
Phys. Lett. B 550 (2002) 47. 

\bibitem{JNH02} D.~Jido, H.~Nagahiro, S.~Hirenzaki, Phys. Rev. C 66 
(2002) 045202. 

\bibitem{FGM13} E.~Friedman, A.~Gal, J.~Mare\v{s}, Phys. Lett. B 725 (2013) 
334. 

\bibitem{CFG14} A.~Ciepl\'{y}, E.~Friedman, A.~Gal, J.~Mare\v{s}, Nucl. Phys. 
A 925 (2014) 126. 

\bibitem{Gal14} A.~Gal, E.~Friedman, N.~Barnea, A.~Ciepl\'{y}, J.~Mare\v{s}, 
D.~Gazda, Acta Phys. Polon. B 45 (2014) 673. 

\bibitem{Mares16} J.~Mare\v{s}, N.~Barnea, A.~Ciepl\'{y}, E.~Friedman, A.~Gal, 
EPJ Web of Conf. 130 (2016) 03006. 

\bibitem{Wilkin16} C.~Wilkin, EPJ Web of Conf. 130 (2016) 01007. 

\bibitem{KWi15} B.~Krusche, C.~Wilkin, Prog. Part. Nucl. Phys. 80 (2015) 43. 

\bibitem{BFG15} N.~Barnea, E.~Friedman, A.~Gal, Phys. Lett. B 747 (2015) 345. 

\bibitem{XLO16} J.J.~Xie, W.H.~Liang, E.~Oset, P.~Moskal, M.~Skurzok, 
C.~Wilkin, Phys. Rev. C 95 (2017) 015202. 

\bibitem{AAB16} P.~Adlarson, et al. (WASA-at-COSY Collaboration), 
Nucl. Phys. A 959 (2017) 102. 

\bibitem{BBFG17} N.~Barnea, B.~Bazak, E.~Friedman, A.~Gal, Phys. Lett. B 771 
(2017) 297. 





\bibitem{Fix17} A.~Fix, O.~Kolesnikov, Phys. Lett. B 772 (2017) 663. 

\bibitem{MNC77} D.R.~Thompson, M.~LeMere, Y.C.~Tang, Nucl. Phys. A 286 
(1977) 53. 

\bibitem{AV402} R.B.~Wiringa, S.C.~Pieper, Phys. Rev. Lett. 89 (2002) 182501. 



\bibitem{GW05} A.M.~Green, S.~Wycech, Phys. Rev. C 71 (2005) 014001. 

\bibitem{CS13} A.~Ciepl\'{y}, J.~Smejkal, Nucl. Phys. A 919 (2013) 46. 

\bibitem{KSW95} N.~Kaiser, P.B.~Siegel, W.~Weise, Phys. Lett. B 362 (1995) 23. 

\bibitem{MBM12} M.~Mai, P.C.~Bruns, U.-G.~Mei{\ss}ner, Phys. Rev. D 86 (2012) 
094033. 

\bibitem{IOV02} T.~Inoue, E.~Oset, M.J.~Vicente Vacas, Phys. Rev. C 65 (2002) 
035204. 

\bibitem{HW08} T.~Hyodo, W.~Weise, Phys. Rev. C 77 (2008) 035204. 

\bibitem{DHW08} A.~Dot\'{e}, T.~Hyodo, W.~Weise, Nucl. Phys. A 804 (2008) 
197. 

\bibitem{DHW09} A.~Dot\'{e}, T.~Hyodo, W.~Weise, Phys. Rev. C 79 (2009) 
014003. 

\bibitem{BGL12} N.~Barnea, A.~Gal, E.Z.~Liverts, Phys. Lett. B 712 (2012) 
132. 

\bibitem{GL81} B.F.~Gibson, D.R.~Lehman, Phys. Rev. C 23 (1981) 404. 

\bibitem{ADN13} I. Angeli, K.P. Marinova, At. Data and Nucl. Data Tables 99 
(2013) 69. 

\bibitem{EK80} J.M.~Eisenberg, D.S.~Koltun, {\it Theory of Meson 
Interactions with Nuclei} (Wiley, New York, 1980) ISBN 0-471-03915-2. 

\bibitem{KMT59} A.K.~Kerman, H.~McManus, R.M.~Thaler, Ann. Phys. 8 (1959) 551, 
reprinted in 281 (2000) 853. 

\bibitem{FGH71} H.~Feshbach, A.~Gal, J.~H\"{u}fner, Ann. Phys. 66 (1971) 20. 

\end{thebibliography}
\end{document}